\documentclass[12pt]{article}

\usepackage{graphicx}
\usepackage{scicite}
\usepackage{float}
\usepackage{enumitem}
\usepackage{gensymb}
\usepackage{amsmath}
\usepackage{xcolor}

\usepackage{times}

\renewcommand{\vec}[1]{\boldsymbol{\mathrm{#1}}}
\newcommand{\mean}[1]{\langle #1 \rangle}

\title{Log it: How to fit an active Brownian particle's mean squared displacement with improved parameter estimation} 
\author
{Maximilian R. Bailey,$^{\ast}$\textit{$^{a}$} Alexander R. Sprenger,\textit{$^{b,c}$}  Fabio Grillo,\textit{$^{a}$}\\ Hartmut Löwen,\textit{$^{b}$} and Lucio Isa$^{\ast}$\textit{$^{a}$}\\
\normalsize{$^{a}$Laboratory for Soft Materials and Interfaces, Department of Materials, ETH Z{\"u}rich,} \\ \normalsize{Vladimir-Prelog-Weg 5, 8093 Z{\"u}rich, Switzerland}\\
\normalsize{$^{b}$Institut f\"ur Theoretische Physik II: Weiche Materie,} \\
\normalsize{Heinrich-Heine-Universit\"at D\"usseldorf, 40225 D\"usseldorf, Germany} \\
\normalsize{$^{c}$ Institut für Physik, Otto-von-Guericke-Universit\"at Magdeburg,} \\
\normalsize{Universit\"atsplatz 2, 39106 Magdeburg, Germany}\\
\normalsize{$^\ast$To whom correspondence should be addressed:} \\
\normalsize{E-mail: maximilian.bailey@mat.ethz.ch;}\\ 
\normalsize{E-mail: lucio.isa@mat.ethz.ch}
}

\date{}

\begin{document} 


\baselineskip24pt


\maketitle

% Keywords: Please provide a minimum of three and a maximum of seven keywords, separated by commas

%\keywords{active Brownian particle, mean squared displacement, regression, fitting, statistical analysis, tutorial}

% Abstract should be written in the present tense and impersonal style (i.e., avoid we), and be at most 200 words long
\begin{abstract}
The active Brownian particle (ABP) model is widely used to describe the dynamics of active matter systems, such as Janus microswimmers. In particular, the analytical expression for an ABP's mean-squared-displacement (MSD) is useful as it provides a means to describe the essential physics of a self-propelled, spherical Brownian particle. However, the truncated or ``short-time'' form of the MSD equation is typically fitted, which can lead to significant problems in parameter estimation. Furthermore, heteroscedasticity and the often statistically dependent observations of an ABP's MSD lead to a situation where standard ordinary least squares (OLS) regression will obtain biased estimates and unreliable confidence intervals. Here, we propose to revert to always fitting the full expression of an ABP's MSD at short timescales, using bootstrapping to construct confidence intervals of the fitted parameters. Additionally, after comparison between different fitting strategies, we propose to extract the physical parameters of an ABP using its mean logarithmic squared displacement (MLSD). These steps improve the estimation of an ABP’s physical properties, and provide more reliable confidence intervals, which are critical in the context of a growing interest in the interactions of microswimmers with confining boundaries and the influence on their motion.
\end{abstract}

\newpage
\vspace{1cm}

%\section{Introduction}

Overdamped active Brownian motion is often invoked to describe the physics of experimental realisations of active matter \cite{Howse2007,Dietrich2017}. The ``active Brownian particle's'' (ABP) motion is described using Langevin dynamics in the overdamped (inertia-free) regime, and consists of an object simultaneously subjected to thermal fluctuations and directed self-propulsion. In this model, the particle moves with a constant velocity $V_{0}$ in the direction of its internal orientation axis $\boldsymbol{\hat u}$, which fluctuates over time due to rotational Brownian motion \cite{TenHagen2011}. Particles therefore travel ballistically over times shorter than the characteristic time scale for rotational diffusion (persistent motion), displaying diffusive motion (with a larger, effective diffusion coefficient) at longer times as their direction of motion is randomized \cite{Bechinger2016}. This model provides meaningful statistical quantities such as an analytical description for the mean-squared-displacement (MSD) of spherical microswimmers, which often shows good agreement with experimental findings \cite{Zheng2013}. Most analyses in the experimental literature on microswimmers are in fact based on parameters estimated by fitting the sample MSD to the ABP model, extracting particle velocity $V_{0}$, translational diffusivity $D_{T}$, and rotational diffusivity $D_{R}$. In two spatial dimensions, the ABP model prescribes the following expression for the MSD $\langle\Delta \vec{r}^{2}(\tau)\rangle$ as a function of lag time $\tau$ \cite{Howse2007,Lowen2020}
\begin{align}
\langle\Delta \vec{r}^{2}(\tau) \rangle \,&=\, 4D_{T}\tau + \frac{2V_0^2}{D_R^2}  \bigg( D_R \tau - 1 + e^{- D_R \tau} \bigg).  \label{eqn:fullMSD}
\end{align}
The standard approach to parameter estimation from a defined model is to use ordinary least squares (OLS) regression \cite{VandenBos2007,Mestre2020} following
\begin{equation}
\vec{\hat\theta} = \operatorname*{argmin}_\theta \sum_{i=1}^{P}(Y_{i}-f_{i,\theta})^{2}, \\
\label{eqn:ls_regress}
\end{equation}
where $\vec{\hat\theta}$ is the vector of estimated parameters, $Y_{i}$ are individual observations from the data set $P$ (here given by the sample MSD after a given lag time $\tau$), and $f_{i,\theta}$ corresponds to the values of the fitted model (here given by the theoretical prediction, see Eq.~\eqref{eqn:fullMSD}). Here, $\operatorname*{argmin}_\theta$ finds the vector $\vec{\theta}$ which minimises the objective function. In practice, there are two main strategies to determine the sample MSD of a population of particles from their coordinates: one can either perform an ensemble average or a time average over the displacements. Ensemble-averaging over many particles preserves the statistical independence of the observations and efficiently averages out spurious noise \cite{Novotny2019}, but collecting sufficient statistics in the dilute limit where Eq.~\eqref{eqn:fullMSD} holds is experimentally challenging.

Therefore, one often typically resorts to the calculation of the MSD via time-averaging the displacements of a few ABP trajectories followed over time. More over, time-averaging is advantageous in that it describes the physics of individual microswimmers, whereas studying the EMSD removes information about the heterogeneities present within the studied system, such as particles displaying atypical motion or changing dynamics within different spatial domains \cite{Kepten2013}. The time-averaged MSD (TAMSD) of a single particle at a lag-time $n\Delta\tau$ is calculated as
\begin{equation}
  \mathrm{TAMSD}\mid_{n\Delta\tau} = \sum_{m=1}^{M-n} \frac{ \big( \vec{r}\big((n+m)\Delta\tau\big) - \vec{r}(m\Delta\tau) \big)^2}{M-n},
  \label{eqn:TAMSD_eqn}
\end{equation}
where ${\vec r}\big((n+m)\Delta\tau\big)$ is the particle position at lag time $n\Delta\tau$ from its previous (reference) position ${\vec r}(m\Delta\tau)$, for a trajectory of length $M$.
By collecting sufficiently long trajectories, there is the implicit assumption that statistically robust averaging is performed, which is required for accurate parameter estimation with OLS. 

However, there are several key assumptions that must be satisfied when using least-squares regression: of these, two can be violated when evaluating the MSD of an ABP. The rotational and time symmetry of a theoretical ABP ensures that consecutive \textit{non-overlapping} squared displacements are statistically independent, but, non-idealities in experimental systems can create hidden correlations, thereby violating the assumption of statistically independent measurements \cite{Fogelmark2018,Qian1991}. Furthermore, to increase the statistics, one typically evaluates \textit{overlapping} squared displacements when investigating ABPs, which are in fact correlated (see below for further discussion). This impacts the reliability of the estimated confidence intervals, which can become unrealistically narrow. In the case of statistically-dependent measurements, confidence intervals for estimated parameters can nevertheless be constructed by fitting the model to bootstrapped datasets from experimental values \cite{Efron1994,Fogelmark2018}.

The second violation is the assumption of homoscedasticity in the error terms of the MSD. There are two sources for heteroscedasticity (non-constant variance) within the error terms of the MSD with lag time. First, as we show later, the theoretical population variance of an ABP's MSD increases with lag time. Furthermore, the number of data points used to estimate the TAMSD decreases with increasing lag time when evaluating single trajectories, further increasing the sampling error. These factors, coupled with the presence of localisation errors at shorter time scales \cite{Kepten2015}, creates a situation where there is an optimal lag time over which the TAMSD of a particle should be evaluated to obtain proper fits of its physical properties \cite{Michalet2010,Devlin2019}.

Weighted least squares (WLS) regression is often implemented in order to reduce the dependence of the fit on data points with greater variance, following
\begin{equation}
\boldsymbol{\hat\theta} = \operatorname*{argmin}_\theta \sum_{i=1}^{P} w_{i,\theta}(Y_{i}-f_{i,\theta})^{2}, \\
\label{eqn:wls_regress_general}
\end{equation}
where $\vec{\hat\theta}$ is again the vector of estimated parameters, $Y_{i}$ are the $P$ data observations, $w_{i,\theta}$ are the weights, and $f_{i,\theta}$ is the model fitted. Here, $\operatorname*{argmin}_\theta$ now finds the vector $\vec{\theta}$ which minimises the weighted objective function. The objective function can be weighted by the inverse of the analytical expression of the population variance (here, the variance of the squared displacements) as an estimation of the sample error of the mean \cite{Michalet2010,Saxton1997}. The variance of the mean of a random variable $X$, i.e. $\text{E}[X] = \sum_{i=1}^{N} X_{i}/N$, can be obtained using the variance sum law for uncorrelated variables as
\begin{equation}
    \text{Var}\big[ \text{E}[X] \big]=\text{Var}\bigg[ \sum_{i=1}^{N}\frac{X_{i}}{N} \bigg] = \frac{1}{N^{2}} \sum_{i=1}^{N} \text{Var}\bigg[ X_{i} \bigg] = \frac{\sigma^{2}}{N}, %  = \frac{1}{N^{2}} N\sigma^{2} 
  \label{eqn:sampling_error}
\end{equation}
where $N$ is the sample size, and $\sigma^2$ is the variance of the random variable $X$. Thus, from Eq.~\eqref{eqn:sampling_error}, we obtain the following expression for the weights $w_{i,\theta}$
\begin{equation}
    w_{i,\theta} = \frac{1}{\text{Var}\big[ \text{E}[X] \big]}= \frac{N_i}{\sigma_{i,\theta}^{2}},
  \label{eqn:weights_explicit}
\end{equation}
where $N_i$ is the number of statistically independent data points contributing to each observation $i$, and $\sigma_{i,\theta}^{2}$ is the population variance of each observation $i$, in terms of the fitted values $\vec{\theta}$. 

Nonetheless, the standard approach in the literature is parameter estimation from TAMSDs using unweighted least-squares regression \cite{Arque2019}. Additionally, perhaps the most widespread expression that is fitted is the so-called ``short-time'' MSD of ABPs \cite{Wang2021b} (Eq.~\eqref{eqn:short_ABP}). First proposed by Howse et al. for the analysis of Janus catalytic microswimmers \cite{Howse2007}, the short-time MSD equation approximates the full MSD (Eq.~\eqref{eqn:fullMSD}) at an arbitrarily short time lag, typically defined as 10\% of the characteristic persistence or rotational diffusion time $\tau_R = 1/ D_{R}$, using a MacLaurin series expansion assuming $\tau/\tau_{R} \rightarrow 0$ \cite{Lowen2020}
\begin{equation}
  \mean{\Delta \vec{r}^{2}(\tau)} \sim 4D_{T}\tau + V_0^{2}\tau^{2} .
  \label{eqn:short_ABP}%\\
\end{equation} 
This simplification provides reasonable fits to the experimental TAMSD of single particles under certain conditions, particularly in relation to the extraction of microswimmer velocities \cite{Howse2007,Arque2019,Pourrahimi2019,Sridhar2020,Ketzetzi2020,Bailey2021b}. However, care should be taken when fitting this truncated form of the MSD to arbitrarily short experimental trajectories, as it can lead to the spurious detection of velocity in the presence of experimental artifacts \cite{Dunderdale2012}. The problems associated with the standard fitting of the truncated form of the MSD were comprehensively demonstrated by Mestre et al \cite{Mestre2020}. Interestingly, their proposed solution was to expand the MacLaurin series to higher polynomial orders. Nonetheless, we are interested in evaluating the fitting of the full ABP's MSD to the ``short-time'' regime, as the approximation is simply that – an approximation of a theoretical model.

In this work, we propose multiple approaches to improve the fitting of the full ABP MSD model. We verify the robustness of our approach by comparing it against the ``standard'' approach of performing unweighted OLS regression on the truncated form of an ABP's MSD at short lag times. We begin by considering the case where $D_T$ and $D_R$ are coupled by the Einstein relation $D_{T} = d_{p}^{2}D_{R}/3$ to avoid the introduction of additional fitting parameters, and thus allow a fair comparison between the standard approach and our proposed alternatives. In the final section of this study, we then treat $D_R$ as an additional free fitting parameter, corresponding to experimental situations where $D_T$ and $D_R$ are often decoupled. We evaluate the different fitting procedures against simulated ABP trajectories using input values representative of experiments. Specifically, in the coupled case, our ABPs are simulated via Langevin dynamics \cite{Callegari2019}, with an active velocity of $V_{0}= 5~\mu$ms\textsuperscript{-1}, and diffusivities $D_{T}$ = 0.2~$\mu$m\textsuperscript{2}s\textsuperscript{-1} and $D_{R}$ = 0.15~rad\textsuperscript{2}s\textsuperscript{-1}. The simulations are numerically solved at 1~ms increments, and sampled at 20 frames per second (fps) for 60~s to replicate experimental videos.

Properly applied, there are several advantages to the standard approach of fitting the TAMSD at short timescales. Generally, the scatter of the sample MSD will increase with lag time. This increase is not only caused by the decrease in data points for a trajectory of a given length, but also due to the growing correlation between sequential observations (see Eq.~\eqref{eqn:TAMSD_eqn}). Therefore, the fitting of the MSD to unnecessarily long lag times is generally discouraged \cite{Saxton1997}. By evaluating the TAMSD over a time period during which the variance does not grow significantly, the effects of heteroscedasticity on parameter estimation are reduced \cite{Michalet2010}. Nonetheless, the term ``short lag times'', where the simplified expression holds, is frequently flawed since it is often arbitrarily defined and used in the literature. Furthermore, by eliminating the opportunity to fit $D_{R}$, the truncated form of Eq.~\eqref{eqn:short_ABP} removes characteristic information on the physics of ABPs. Finally, for smaller particles, the characteristic persistence time may be so short that only a few data points can be used to fit the expression, unless experiments are performed at very high frame rates, introducing measurement error and reducing the accuracy of parameter estimation \cite{Kepten2015}. 

\begin{figure}
\centering
  \includegraphics[width=0.9\linewidth]{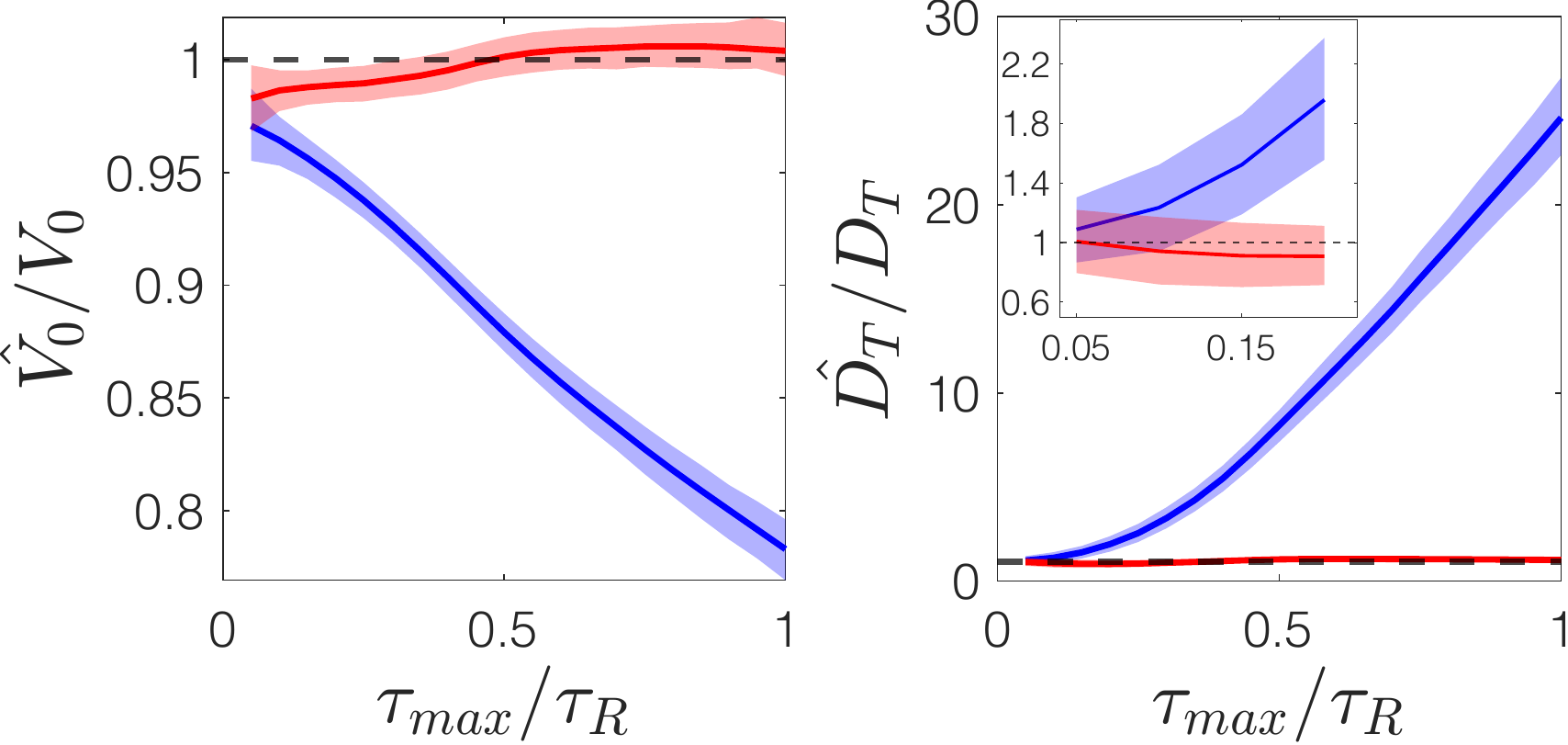}
  \caption{Parameter estimation from fitting the truncated (blue) and full MSD (red) expression to simulated data (estimates $\hat{V_{0}}, \hat{D_{T}}$ respectively normalised to the simulation inputs $V_{0,sim}, D_{T,sim}$). The same trajectory is fitted to increasing maximal lag times $\tau_{max}$, up to the persistence time of an ABP ($\tau_R$). 95\% confidence intervals are obtained by bootstrapping. Inset (right): Fits of  $D_{T}$ for short $\tau_{max}$, indicating the rapid deviation from the input simulation value when using the truncated expression.}
  \label{fig:Fig1}
\end{figure}

There are, in fact, further model-specific problems associated with fitting the truncated form of the MSD. As seen in Eq.~\eqref{eqn:ls_regress}, OLS regression is weighted towards larger values, i.e., MSD values at longer lag times. If left untreated, the fitting of the MSD will therefore be weighted towards the ``long-time diffusive'' regime of the ABP \cite{Bechinger2016}. Moreover, due to the monotonically growing variance in the error terms of the MSD (discussed below in more detail), this procedure assigns greater importance to more uncertain values, leading to poorer estimates. The effects of these considerations are illustrated by comparing the estimates for $D_{T}$ and $V_{0}$ obtained by fitting the truncated and full-form of the MSD equation to simulated trajectories (see Fig. \ref{fig:Fig1}). 

The problems of using Eq.~\eqref{eqn:short_ABP} become quickly apparent as the lag times evaluated increase beyond small fractions of the characteristic relaxation time $\tau_R$. As the estimated velocity decreases, the fitted $D_{T}$ value rapidly increases to over an order of magnitude greater than the simulation input (see Fig. \ref{fig:Fig1}, blue). The inverse relationship between $V_0$ and $D_T$ can be understood by their respective contributions to the overall MSD of an ABP. The increasingly diffusive nature of an ABP's motion with time \cite{Bechinger2016} results in an over-estimated $D_T$ at the expense of a reduction in the fitted $V_0$. This problem is caused by the absence of the $D_R$-related terms in Eq.~\eqref{eqn:short_ABP}, which would otherwise result in the crossover to a long-time diffusive regime (see Eq.~\eqref{eqn:fullMSD}). In short, due to the systematic errors associated with using Eq.~\eqref{eqn:short_ABP} we strongly advise against its use when fitting the MSD of ABPs. 

In contrast, the bootstrapped confidence intervals of the estimated parameters using Eq.~\eqref{eqn:fullMSD} more often include the true simulation input values for different maximal lag times $\tau_{max}$ and also converge to reasonable values as the lag time evaluated approaches the characteristic rotational relaxation time $\tau_{R}$ (see Fig. \ref{fig:Fig1}, red). Fitting Eq.~\eqref{eqn:fullMSD} also carries the advantage of not assuming a limited short-time regime, enabling the fitting to longer lag times and thus providing more data points for better parameter estimation. We again emphasize that we do not fit $D_R$ as a free parameter here but instead assume that the Einstein relation $D_{T} = d_{p}^{2}D_{R}/3$ holds and fit Eq.~\eqref{eqn:fullMSD} accordingly. However, decoupling $D_T$ and $D_R$ better approximates experimental situations where the presence of confining boundaries \cite{Goldman1967}, activity \cite{Ebbens2010} (both \cite{Das2015,Simmchen2016}), or external fields \cite{Sprenger2020} can have a different effect on rotation and translation respectively.

\begin{figure}
\centering
  \includegraphics[width=0.9\linewidth]{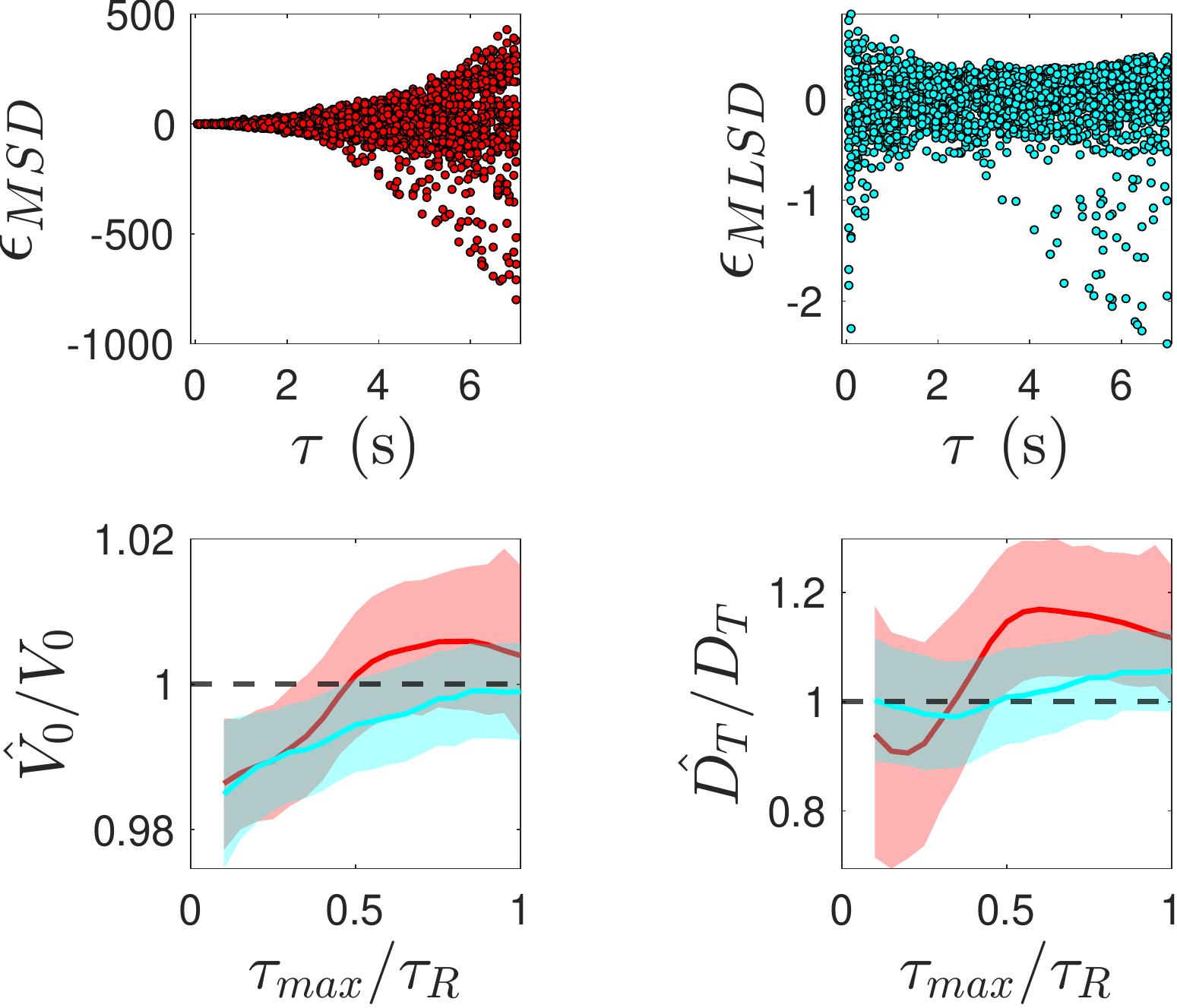}
  \caption{Top row: Plots of the residuals ($\epsilon_{MSD}$) from the MSD based on the point estimates in the bottom row for $\tau_{max}/\tau_{R} = 1$ as a function of lag time $\tau$. Left (red): Residuals of Eq.~\eqref{eqn:fullMSD} fitted to unprocessed data. Right (cyan): Residuals of log(Eq.~\eqref{eqn:fullMSD}) fitted to log-transformed data. The extent of heteroscedasticity is clearly reduced, as the variance remains relatively constant with $\tau$ after log transformation. Bottom row: Parameter estimation without (red) and with (cyan) log-transformation of the data and the model.}
  \label{fig:Fig2}
\end{figure}

Despite the significant improvement in estimating the physical parameters of an ABP by using the full form of its MSD equation, this operation still does not address underlying statistical issues such as heteroscedasticity of the data. The presence of heteroscedasticity can be clearly observed in the residuals of the fitted ABP model (see Fig. \ref{fig:Fig2}, top row, red). One of the most frequently used heuristic approach to address heteroscedasticity is to log-transform the data and fit the model's log-transformed analog. Log transforms work particularly well for right skew, constantly positive, and increasing data, such as the case for the ABP's MSD. Studying the ``mean logarithmic squared displacement'' (MLSD) has previously been suggested to improve the estimation of the distribution of anomalous diffusion coefficients in a population of heterogeneous particles \cite{Kepten2013}. 

By fitting the log-transformed (cyan) data, we observe a clear reduction of the heteroscedasticity of the residuals. This provides improved estimated fits and confidence intervals obtained from bootstrapping. In Fig. \ref{fig:Fig2} (bottom row), we highlight the improvement in fitting after this simple pre-processing step, evaluating the same trajectory as in Fig. \ref{fig:Fig1} but now with the log-transformed, full MSD ABP fit included as a comparison to the full fit without log-transformation. We see both a reduction in the width of the confidence intervals and a smaller difference between the point estimate and the input simulation values. In particular, the estimates for $D_T$ are notably improved.

As a next step, we turn to WLS regression as a tool for determining the parameters of an ABP. As previously alluded to, within the WLS regression approach, one typically relates the weights to the variance of the expectation value (see Eq.~\eqref{eqn:weights_explicit}). Under the assumption that all observations are statistically independent, the variance of the expectation value can be obtained from the population variance itself, using the variance sum law as shown in Eq.~\eqref{eqn:sampling_error}. In the remainder of this work, we will follow this approach and specify the weights in terms of the theoretical result for the variance of the mean squared displacement \cite{Zheng2013,Kurzthaler2017,Sevilla2021}
\begin{align}
	\sigma^{2}(\tau) =&   \mean{\Delta \vec{r}^{4}(\tau)} - \mean{\Delta \vec{r}^{2}(\tau)}^{2} \nonumber\\ 
	=& 16 D_T^2 \tau^2 + 16 D_T \tau \frac{V_0^2}{D_R^2}  \bigg( D_R \tau - 1 + e^{-D_R \tau} \bigg)    \nonumber \\
	&+ \frac{V_0^4}{D_R^4}   \bigg( 4 D_R^2 \tau^2 - 22 D_R \tau + \frac{79}{2}- \frac{64}{3} D_R \tau e^{-D_R \tau}    \nonumber \\
	& - \frac{320}{9} e^{-D_R \tau} - 4 e^{-2D_R \tau} + \frac{1}{18} e^{-4D_R \tau}  \bigg).
	\label{eqn:MSD_variance}
\end{align}
We note that this result is only an exact representation of the variance of the mean if non-overlapping squared displacements are considered. For overlapping displacements, a proper analysis requires additional covariance contributions in Eq.~\eqref{eqn:sampling_error}, describing the correlation between subsequent displacements. In that case, we will still employ Eq.~\eqref{eqn:MSD_variance}, however, as an approximation. Equipped with this expression, we can now investigate the presence of heteroscedasticity in an ABP's MSD, and attempt to minimize its effects on parameter estimation using WLS regression. Last, as discussed before, we stress that in an experimental context, there might be further hidden correlations between square displacements requiring special consideration, whose evaluation lies beyond the aims of this work. 
\begin{figure}
\centering
  \includegraphics[width=0.9\linewidth]{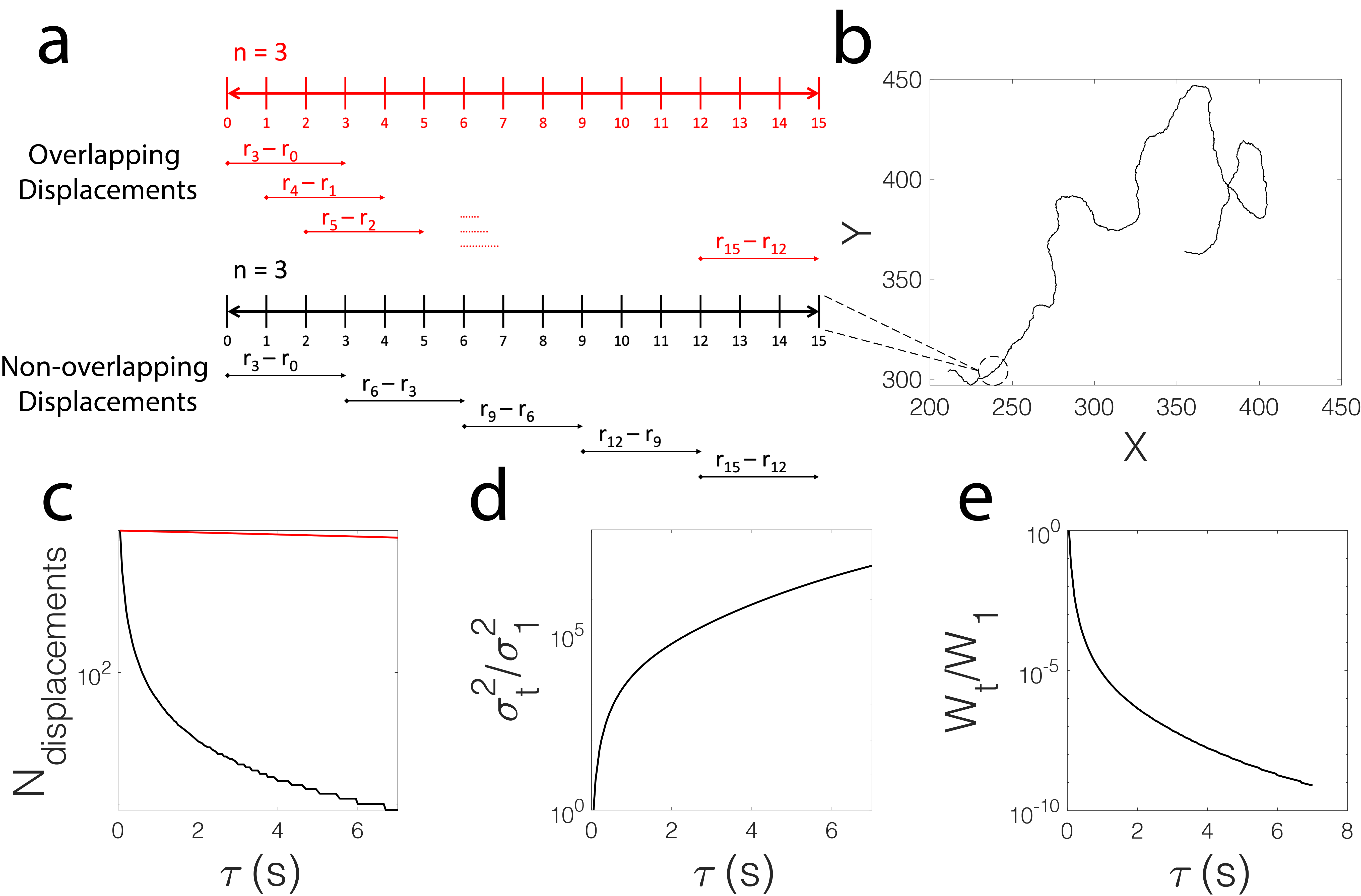}
  \caption{a) Definition of overlapping and non-overlapping displacements from the simulated trajectory shown in b). c) Number of displacements as a function of lag time when overlapping (red) and non-overlapping (black) displacements are evaluated. d) Normalized variance of the MSD as a function of $\tau$ ($\sigma_1$ is the variance at the shortest lag time $\tau_1$ = 0.05~s), as derived by Equation \eqref{eqn:MSD_variance}. e) Corresponding normalized weight at time $\tau$ ($w_1$ is the weight at $\tau_1$ = 0.05~s) extracted according to Eq.~\eqref{eqn:wls_regress_general} as a function of $\tau$ for the TAMSD of a single particle.}
  \label{fig:Fig4} 
\end{figure}
As alluded to above, the TAMSD of particles can be evaluated with one of two different approaches: by determining the overlapping or non-overlapping particle displacements (see Fig. \ref{fig:Fig4} a,b). Evaluating non-overlapping squared displacements reduces the correlation between subsequent observations of motion in experimental scenarios and removes it entirely within the framework of the ABP model. However, in this case, the decay in the number of displacements is hyperbolic, decreasing much more rapidly than when overlapping displacements are evaluated (see Fig. \ref{fig:Fig4} c). Furthermore, only using non-overlapping displacements leads to a different sampling of points along the trajectory depending on how many prime factors are present in the number of the time step. These factors lead to a situation where using overlapping displacements typically improves fitting performance and is generally preferable \cite{Saxton1997}. 

%\section{Results \& Discussion}

We now discuss the potential benefits of applying the weighting coefficient to minimize the effects of the large and high-variance long lag time values in the objective function (see Eq.~\eqref{eqn:wls_regress_general}). From Eq.~\eqref{eqn:MSD_variance}, we find that the variance increases with time (see Fig. \ref{fig:Fig4} d), and combined with the decay in the number of observations (see Fig. \ref{fig:Fig4} c), we obtain with Eq.~\eqref{eqn:sampling_error} a weighting vector that rapidly decays with time (see Fig. \ref{fig:Fig4} e). This in turn demonstrates that the low numbers of observations at longer time scales, which inherently have a larger variance due to the nature of the TAMSD, will have a significantly reduced influence on parameter estimation.

\begin{figure}
\centering
  \includegraphics[width=0.9\linewidth]{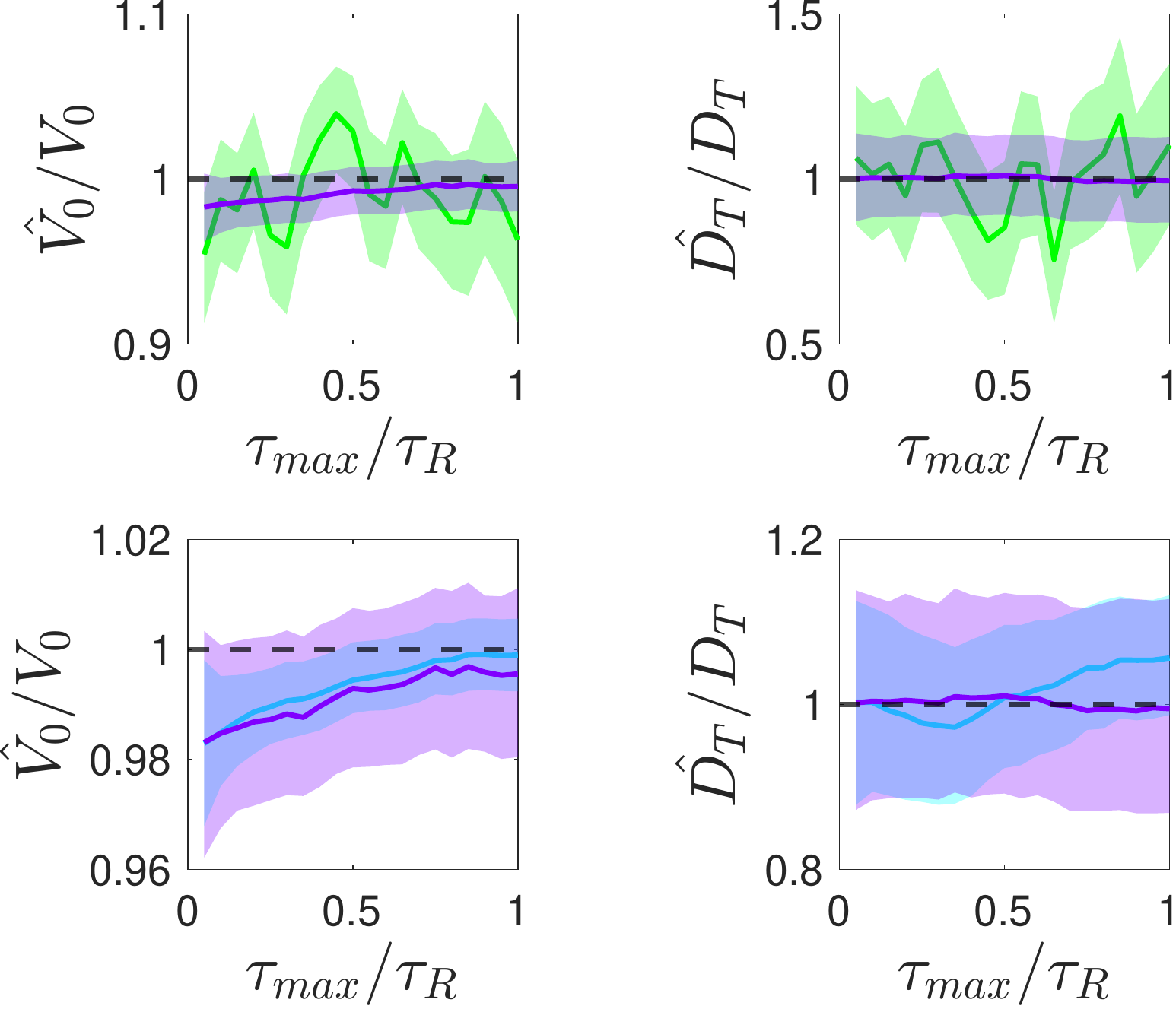}
  \caption{Top row: Parameter estimation using WLS regression on non-overlapping (green) and overlapping (purple) displacements. Bottom row: Parameter estimation on overlapping displacements using WLS regression (purple) and the MLSD (cyan)}
  \label{fig:Fig5}
\end{figure}

We now fit the TAMSD of a single particle using WLS regression, beginning with the analysis of non-overlapping displacements (see Fig. \ref{fig:Fig5}, top row, green). We note a significant instability in the point estimates and confidence intervals, particularly for $\hat{D_T}$, in direct comparison to the fits obtained with the MLSD. Therefore, we also evaluate the performance of WLS regression on overlapping displacements, noting that the underlying assumption of statistically independent observations no longer holds (see Fig. \ref{fig:Fig5}, top row, purple). Comparing the overlapping to the non-overlapping case, we find that the resulting confidence intervals and point estimates for WLS regression are much narrower and less subject to fluctuations. We expect this discrepancy arises, in large part, from the statistical issues associated with evaluating non-overlapping displacements, as described in \cite{Saxton1997}. Motivated by the improved parameter estimation, we continue to evaluate WLS regression using overlapping displacements for the rest of this work.

We now compare the performance of the MLSD and WLS regression for parameter estimation from overlapping displacements (see Fig. \ref{fig:Fig5}, bottom row). Although the resulting confidence intervals are broader for the WLS regression than for the MLSD, we note that in the former case the estimate for $D_{T}$ is more stable, and the true simulation input parameters are included for all values of $\tau_{max}$. We conclude that for a 2-parameter fit, where $D_{T} = d_{p}^{2}D_{R}/3$, the estimates obtained from WLS regression and OLS regression of the MLSD are similar. 

So far, we have only considered particles satisfying the ideal condition where $D_T$ and $D_R$ are related by the Einstein relationship for freely diffusing spherical particles. However, in many situations, e.g. when in proximity with a solid wall, $D_T$ and $D_R$ are likely to be decoupled \cite{Goldman1967,Ebbens2010,Das2015,Simmchen2016,Dietrich2017}, and it is therefore important, in most experimental realizations of ABPs, to fit these parameters separately. We  account for these circumstances by modifying the value of $D_T$, while keeping the same value of $D_R$ in our simulations. In particular, we modify the translational diffusivity by applying Faxen's correction factor to $D_T$, as if to mimic the presence of a solid wall 250 nm away from the particle surface  \cite{Ketzetzi2020a}. This correction approximately reduces the theoretical $D_T$ value we initially used by half.

\begin{figure}
\centering
  \includegraphics[width=0.9\linewidth]{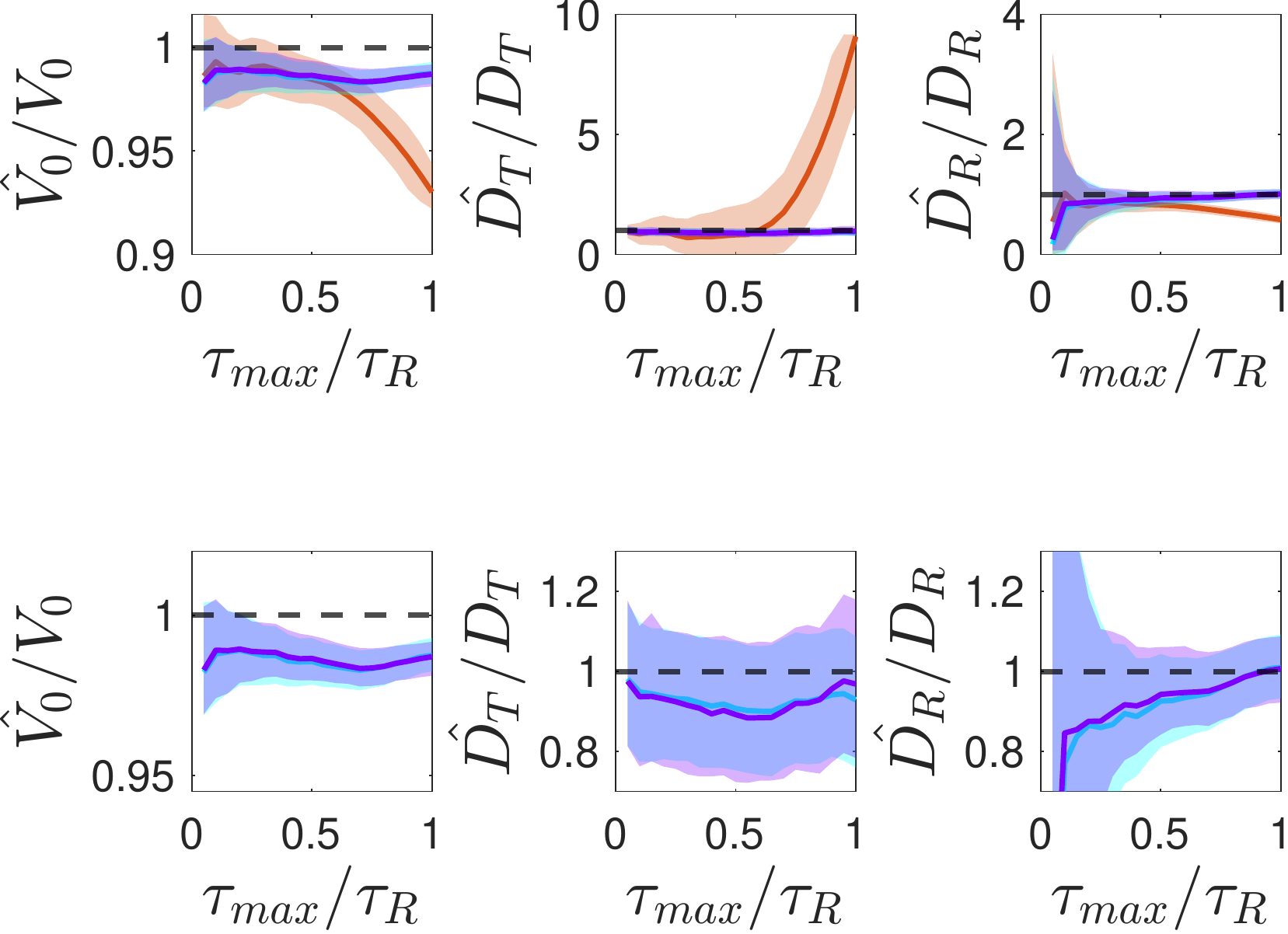}
  \caption{Parameter estimation of an ABP's MSD where $D_T$ and $D_R$ are uncoupled, using WLS regression (purple), MLSD (cyan), and the 3rd order truncation of the MSD equation (orange). Top row: Comparison of the 3 fitting approaches. The truncated expression clearly performs worse, particularly at larger $\tau_{max}$ (see the estimates for $D_T$ and $D_R$). Bottom row: Only the MLSD and WLS regression are represented for better visualisation. }
  \label{fig:Fig6}
\end{figure}

In Fig. \ref{fig:Fig6}, we compare the performance of the MLSD and WLS regression approaches when estimating the parameters $V_0$, $D_T$, and $D_R$ (blue and purple, respectively). We also study the truncated MSD equation expanded to third order, as outlined in \cite{Mestre2020} (Fig. \ref{fig:Fig6}, top row, orange). This expression is obtained by evaluating the MacLaurin series expansion of Eq.~\eqref{eqn:fullMSD} to the 3rd order
\begin{equation}
  \mean{\Delta \vec{r}^{2}(\tau)} \sim 4D_{T}\tau + V_0^{2}\tau^{2} - \frac{V_0^2}{3\tau_R}\tau^3.
  \label{eqn:short_ABP_3rdorder}%\\
\end{equation} 
We find that as before, the truncated form of the full MSD equation is not able to satisfactorily capture the input simulation parameters, an effect which is particularly noticeable for $D_T$ as $\tau_{max}$ increases, as previously observed in Fig. \ref{fig:Fig1}.

When evaluating overlapping displacements using WLS regression and the MLSD, we note a remarkable overlap in both the point estimates and confidence intervals obtained at $\tau_{max} \geq 0.35$ (see Fig. \ref{fig:Fig6}, bottom row). This observation indicates that both the log-transformation and weighting of the data have a similar effect on addressing the heteroscedasticity present in an ABP's MSD. In both instances, we also note the instability of the short-time estimates for $\hat{D_R}$, which is unsurprising given the independence of the MSD from $D_R$ at short lag times (see Eq.~\eqref{eqn:short_ABP}) and considering the previous results in Fig. \ref{fig:Fig1}.

In conclusion, the ABP model provides a useful framework for studying  the motion of microswimmers and extracting meaningful physical properties from mean quantities. However, prudence is necessary when estimating these parameters from a "blind" fitting of MSDs since hidden correlations may arise in experimental systems. Therefore, we recommend constructing confidence intervals by bootstrapping in almost all experimental situations. We advise against the use of the truncated form of the MSD equation in all situations. Further steps beyond fitting to short lag times should also be taken to treat the heteroscedasticity of an ABP's MSD. In particular, we find that log-transforming the data before fitting the MLSD equation performs remarkably well in comparison to the standard approaches used in literature, and even gives similar estimates as WLS regression using the theoretical variance of an ABP's MSD. One of the key advantages of this approach is that overlapping displacements can be evaluated, significantly increasing the amount of data available. Furthermore, the simplicity of this approach (simply log-transforming the model and data and fitting to shorter lag times) should assist in its widespread uptake. However, we stress that we have focussed on simulated data, ignoring the effects of deviations from an ideal, spherical ABP \cite{Ebbens2010,Wang2017a,Kummel2013}, the effect of ABP speed on the coupling between $D_T$ and $V_0$ \cite{Tang2018}, and experimental errors from static and dynamic localisation errors \cite{Savin2005,Michalet2010,Kepten2013,Devlin2019}. These are nevertheless critical factors which should be considered when designing experiments and analyzing data. 

\newpage
% Acknowledgements
\medskip
\textit{Acknowledgements} - The authors thank Dr. S. Ketzetzi and Dr. B. ten Hagen for various insightful discussions.
 
\medskip
\textit{Author Contribution Statement} - Author contributions are defined based on the CRediT (Contributor Roles Taxonomy). Conceptualization: M.R.B., F.G. Formal Analysis: M.R.B., A.R.S. Funding acquisition: L.I. Investigation: M.R.B., A.R.S. Methodology: M.R.B., F.G. Software: M.R.B. Supervision: F.G., L.I., H.L. Validation: M.R.B., A.R.S. Visualization: M.R.B. Writing - original draft: M.R.B., A.R.S., L.I. Writing - review and editing: M.R.B., A.R.S., H.L, L.I.      

\newpage
\bibliographystyle{MSP}
\bibliography{Shorttime_MSD}

\begin{thebibliography}{10}
\providecommand{\url}[1]{\texttt{#1}}
\providecommand{\urlprefix}{URL }

\bibitem{Howse2007}
J.~R. Howse, R.~A. Jones, A.~J. Ryan, T.~Gough, R.~Vafabakhsh, R.~Golestanian,
\newblock \emph{Physical Review Letters} \textbf{2007}, \emph{99}, 4 8.

\bibitem{Dietrich2017}
K.~Dietrich, D.~Renggli, M.~Zanini, G.~Volpe, I.~Buttinoni, L.~Isa,
\newblock \emph{New Journal of Physics} \textbf{2017}, \emph{19}, 6.

\bibitem{TenHagen2011}
B.~ten Hagen, S.~{Van Teeffelen}, H.~L{\"{o}}wen,
\newblock \emph{Journal of Physics Condensed Matter} \textbf{2011}, \emph{23},
  19.

\bibitem{Bechinger2016}
C.~Bechinger, R.~{Di Leonardo}, H.~L{\"{o}}wen, C.~Reichhardt, G.~Volpe,
  G.~Volpe,
\newblock \emph{Reviews of Modern Physics} \textbf{2016}, \emph{88}, 4.

\bibitem{Zheng2013}
X.~Zheng, B.~ten Hagen, A.~Kaiser, M.~Wu, H.~Cui, Z.~Silber-Li, H.~L{\"{o}}wen,
\newblock \emph{Physical Review E} \textbf{2013}, \emph{88}, 3 1.

\bibitem{Lowen2020}
H.~L{\"{o}}wen,
\newblock \emph{The Journal of Chemical Physics} \textbf{2020}, \emph{152}, 4
  40901.

\bibitem{VandenBos2007}
A.~van~den Bos,
\newblock \emph{{Parameter Estimation for Scientists and Engineers}},
\newblock John Wiley {\&} Sons, Inc., \textbf{2007}.

\bibitem{Mestre2020}
R.~Mestre, L.~S. Palacios, A.~Miguel-L{\'{o}}pez, X.~Arqu{\'{e}},
  I.~Pagonabarraga, S.~S{\'{a}}nchez,
\newblock \emph{arXiv} \textbf{2020}.

\bibitem{Novotny2019}
F.~Novotn{\'{y}}, M.~Pumera,
\newblock \emph{Scientific Reports} \textbf{2019}, \emph{9}, 1 1.

\bibitem{Kepten2013}
E.~Kepten, I.~Bronshtein, Y.~Garini,
\newblock \emph{Physical Review E} \textbf{2013}, \emph{87}, 5 1.

\bibitem{Fogelmark2018}
K.~Fogelmark, M.~A. Lomholt, A.~Irb{\"{a}}ck, T.~Ambj{\"{o}}rnsson,
\newblock \emph{Scientific Reports} \textbf{2018}, \emph{8}, 1 1.

\bibitem{Qian1991}
H.~Qian, M.~P. Sheetz, E.~L. Elson,
\newblock \emph{Biophysical Journal} \textbf{1991}, \emph{60}, 4 910.

\bibitem{Efron1994}
B.~Efron, R.~J. Tibshirani,
\newblock \emph{{An Introduction to the Bootstrap}},
\newblock Chapman {\&} Hall/CRC, Philadelphia, PA, \textbf{1994}.

\bibitem{Kepten2015}
E.~Kepten, A.~Weron, G.~Sikora, K.~Burnecki, Y.~Garini,
\newblock \emph{PLoS ONE} \textbf{2015}, \emph{10}, 2 1.

\bibitem{Michalet2010}
X.~Michalet,
\newblock \emph{Physical Review E} \textbf{2010}, \emph{82}, 4 1.

\bibitem{Devlin2019}
J.~Devlin, D.~Husmeier, J.~A. Mackenzie,
\newblock \emph{Physical Review E} \textbf{2019}, \emph{100}, 2 1.

\bibitem{Saxton1997}
M.~J. Saxton,
\newblock \emph{Biophysical Journal} \textbf{1997}, \emph{72}, 4 1744.

\bibitem{Arque2019}
X.~Arqu{\'{e}}, A.~Romero-Rivera, F.~Feixas, T.~Pati{\~{n}}o, S.~Osuna,
  S.~S{\'{a}}nchez,
\newblock \emph{Nature Communications} \textbf{2019}, \emph{10}, 1 1.

\bibitem{Wang2021b}
W.~Wang, T.~E. Mallouk,
\newblock \emph{ACS Nano} \textbf{2021}, \emph{15}, 10 15446.

\bibitem{Pourrahimi2019}
A.~M. Pourrahimi, K.~Villa, C.~L. {Manzanares Palenzuela}, Y.~Ying, Z.~Sofer,
  M.~Pumera,
\newblock \emph{Advanced Functional Materials} \textbf{2019}, \emph{29}, 22 1.

\bibitem{Sridhar2020}
V.~Sridhar, F.~Podjaski, J.~Kr{\"{o}}ger, A.~Jim{\'{e}}nez-Solano, B.~W. Park,
  B.~V. Lotsch, M.~Sitti,
\newblock \emph{Proceedings of the National Academy of Sciences of the United
  States of America} \textbf{2020}, \emph{117}, 40 24748.

\bibitem{Ketzetzi2020}
S.~Ketzetzi, J.~{De Graaf}, R.~P. Doherty, D.~J. Kraft,
\newblock \emph{Physical Review Letters} \textbf{2020}, \emph{124}, 4 48002.

\bibitem{Bailey2021b}
M.~R. Bailey, N.~Reichholf, A.~Flechsig, F.~Grillo, L.~Isa,
\newblock \emph{Particle and Particle Systems Characterization} \textbf{2021},
  \emph{2100200} 1.

\bibitem{Dunderdale2012}
G.~Dunderdale, S.~Ebbens, P.~Fairclough, J.~Howse,
\newblock \emph{Langmuir} \textbf{2012}, \emph{28}, 30 10997.

\bibitem{Callegari2019}
A.~Callegari, G.~Volpe,
\newblock \emph{Flowing Matter} \textbf{2019}, \emph{1} 211.

\bibitem{Goldman1967}
A.~J. Goldman, R.~G. Cox, H.~Brenner,
\newblock \emph{Chemical Engineering Science} \textbf{1967}, \emph{22}, 4 637.

\bibitem{Ebbens2010}
S.~Ebbens, R.~A.~L. Jones, A.~J. Ryan, R.~Golestanian, J.~R. Howse,
\newblock \emph{Physical Review E} \textbf{2010}, \emph{82}, 1 15304.

\bibitem{Das2015}
S.~Das, A.~Garg, A.~I. Campbell, J.~Howse, A.~Sen, D.~Velegol, R.~Golestanian,
  S.~J. Ebbens,
\newblock \emph{Nature Communications} \textbf{2015}, \emph{6}, 1 8999.

\bibitem{Simmchen2016}
J.~Simmchen, J.~Katuri, W.~E. Uspal, M.~N. Popescu, M.~Tasinkevych,
  S.~S{\'{a}}nchez,
\newblock \emph{Nature Communications} \textbf{2016}, \emph{7}, 1 10598.

\bibitem{Sprenger2020}
A.~R. Sprenger, M.~A. Fernandez-Rodriguez, L.~Alvarez, L.~Isa, R.~Wittkowski,
  H.~L{\"o}wen,
\newblock \emph{Langmuir} \textbf{2020}, \emph{36}, 25 7066.

\bibitem{Kurzthaler2017}
C.~{Kurzthaler}, T.~{Franosch},
\newblock \emph{Soft Matter} \textbf{2017}, \emph{13}, 37 6396.

\bibitem{Sevilla2021}
F.~J. Sevilla, P.~Castro-Villarreal,
\newblock \emph{Physical Review E} \textbf{2021}, \emph{104}, 6 64601.

\bibitem{Ketzetzi2020a}
S.~Ketzetzi, J.~{De Graaf}, D.~J. Kraft,
\newblock \emph{Physical Review Letters} \textbf{2020}, \emph{125}, 23 238001.

\bibitem{Wang2017a}
X.~Wang, M.~In, C.~Blanc, A.~W{\"{u}}rger, M.~Nobili, A.~Stocco,
\newblock \emph{Langmuir} \textbf{2017}, \emph{33}, 48 13766.

\bibitem{Kummel2013}
F.~K{\"{u}}mmel, B.~{Ten Hagen}, R.~Wittkowski, I.~Buttinoni, R.~Eichhorn,
  G.~Volpe, H.~L{\"{o}}wen, C.~Bechinger,
\newblock \emph{Physical Review Letters} \textbf{2013}, \emph{110}, 19 1.

\bibitem{Tang2018}
E.~M. Tang, P.~T. Underhill,
\newblock \emph{Langmuir} \textbf{2018}, \emph{34}, 36 10694.

\bibitem{Savin2005}
T.~Savin, P.~S. Doyle,
\newblock \emph{Biophysical Journal} \textbf{2005}, \emph{88}, 1 623.

\end{thebibliography}

\end{document}